# On multiplicities in length spectra of arithmetic hyperbolic three-orbifolds

by

J. Marklof*

II. Institut für Theoretische Physik, Universität Hamburg

Luruper Chaussee 149, 22761 Hamburg

Federal Republic of Germany

*Supported by
Deutsche Forschungsgemeinschaft under Contract No. DFG-Ste 241/7-1.
Present address:
Department of Mathematics, Princeton University, Fine Hall - Washington Road, Princeton NJ 08544, USA

**Abstract.** Asymptotic laws for mean multiplicities of lengths of closed geodesics in arithmetic hyperbolic three-orbifolds are derived. The sharpest results are obtained for non-compact orbifolds associated with the Bianchi groups $SL(2, \mathfrak{o}_K)$ and some congruence subgroups. Similar results hold for cocompact arithmetic quaternion groups, if a conjecture on the number of gaps in their length spectra is true. The results related to the groups above give asymptotic lower bounds for the mean multiplicities in length spectra of arbitrary arithmetic hyperbolic three-orbifolds. The investigation of these multiplicities is motivated by their sensitive effect on the eigenvalue spectrum of the Laplace-Beltrami operator on a hyperbolic orbifold, which may be interpreted as the Hamiltonian of a three-dimensional quantum system being strongly chaotic in the classical limit.

# I Introduction

It is well known that there is a deep connection between the geometric properties of an $n$-orbifold (i.e., a space looking locally like $\mathbb{R}^n$ modulo the action of a discrete subgroup of $O(n)$) and the eigenvalue spectrum of its Laplace-Beltrami operator. According to Selberg's theory [1], for finite hyperbolic orbifolds the spectrum is completely determined by its volume, boundary, conic singularities, and – what is most important – by its closed geodesics. Their number up to a given length $l$ grows exponentially [2]:

$$\mathcal{N}(l) \sim \text{Ei}(\tau l) \sim \frac{e^{\tau l}}{\tau l}, \qquad l \to \infty, \tag{1}$$

with the so-called *topological entropy* $\tau$, which is $\tau = n - 1$ for hyperbolic $n$-orbifolds. The similarity with the famous *Prime Number Theorem* "the number of primes below $x$ is asymptotically given by $x/\log x$" is not by chance, and therefore (1) is often denoted as *Prime Geodesic Theorem*.

The exponential proliferation (1) is typical for chaotic classical Hamiltonian systems [3]. Indeed, the geodesic flow on the unit cotangent bundle of a hyperbolic orbifold reflects the properties of an Anosov system [4], which is one of the strongest characterizations of classical chaos.

Concerning the quantum theory of general strongly chaotic systems, similar trace formulae as in Selberg's theory are valid, although only in the semiclassical limit. Such trace formulae were first discovered by Gutzwiller, see [3] for references. The free motion in a hyperbolic orbifold can be viewed as a model for strongly chaotic systems, in the sense that rigorous results allow to explain quantum effects which are as well observed in more realistic settings. Among all hyperbolic orbifolds the arithmetic spaces are the ones for which most results have been achieved, mainly in two, and recently also in three dimensions [5, 6]. On the other hand, arithmetic hyperbolic orbifolds possess a feature which distinguishes them from all other hyperbolic orbifolds, and from all other chaotic systems investigated up to now: It is the strong exponential growth of the mean multiplicities in the length spectrum (i.e., the number of closed geodesics of the same length, averaged over an interval $\Delta l$), which is at least

$$(2) \qquad \langle g(l) \rangle \sim c \, \frac{e^{l/2}}{l}, \qquad l \to \infty,$$

for arithmetic hyperbolic two-orbifolds, compare [7, 8, 9, 10, 11], and

$$(3) \qquad \langle g(l) \rangle \sim c \, \frac{e^{l}}{l}, \qquad l \to \infty,$$

for arithmetic hyperbolic three-orbifolds. $c$ is a constant depending on the considered orbifold. We will discuss in detail, for which kind of orbifolds relation (3) can be proved or conjectured to be valid not just "at least" but exactly, and how (3) serves as an asymptotic lower bound for any other arithmetic hyperbolic three-orbifold. It is interesting to note that there are non-arithmetic orbifolds, which show a similar (but weaker) exponential increase of the multiplicities, e.g., the tetrahedral orbifold $T_8$, where

$$(4) \qquad \langle g(l) \rangle \sim c_1 \, \frac{e^{c_2 l}}{l}, \qquad l \to \infty,$$

with $c_1 \approx 1.410$ and $c_2 \approx 0.643 < 1$, which is a numerical result [12].

As a consequence of the strong increase of the mean multiplicities (2), (3) one is able to explain the attraction of neighbouring eigenvalues of the Laplace-Beltrami operator, which contradicts the widely accepted belief that, for quantum systems being chaotic in the classical limit, the energy spectrum shows eigenvalue repulsion. To be more precise, Bolte [13] has carried out a semiclassical analysis which shows that the short range correlations of the eigenvalue spectrum of arithmetic surfaces tend to those of a Poisson distribution, usually expected for integrable systems [14]. Starting with (3), the results of [13] can be readily extended to three dimensions such that again level attraction is predicted [15]. However, certain assumptions on the characters associated with closed geodesics have to be made within the analysis in both two and three dimensions. The importance of these characters is illustrated in [16], where a two-dimensional hyperbolic billiard is presented, whose quantal spectrum shows level attraction or level repulsion, depending on the choice of boundary conditions.

Along with the third dimension there comes an additional phase angle $\phi$ related to so-called loxodromic orbits, which describes the twist of the neighbourhood of an orbit after one traversal. The notation of a complex length $\ell := l + \mathrm{i}\phi$ will be introduced, where $l$ is the usual (real) length. The mean multiplicity $\langle g(l) \rangle$ is asymptotically connected with the number of closed geodesics (1) and the number $\mathcal{N}^r(l)$ of distinct real lengths of closed geodesics via

$$(5) \qquad \langle g(l) \rangle \sim \frac{d\mathcal{N}(l)}{dl} \left( \frac{d\mathcal{N}^r(l)}{dl} \right)^{-1}, \qquad l \to \infty.$$

In section IV we firstly calculate the asymptotics of the counting function $\mathcal{N}^c(l)$ of distinct complex lengths of closed geodesics for hyperbolic orbifolds whose fundamental group $\Gamma = \mathcal{O}^1$ consists of all units of an quaternion order $\mathcal{O}$ with norm one. We call such groups *arithmetic quaternion groups* for brevity (for definitions see sections II and III). If the set of traces $\operatorname{tr} \Gamma$ of elements of $\Gamma$ is invariant under complex conjugation, then

$$(6) \qquad \mathcal{N}^r(l) \sim \frac{1}{2} \mathcal{N}^c(l), \qquad l \to \infty,$$

see section V. Groups which satisfy that condition are for example orientation-preserving subgroups of reflection groups, i.e., groups generated by the reflections at the faces of a polyhedron. The sharpest results are obtained for the Bianchi groups $\mathrm{SL}(2, \mathfrak{o}_K)$, where $\mathfrak{o}_K$ is the ring of integers of an imaginary quadratic number field $K = \mathbb{Q}(\sqrt{-D})$, $D$ a square-free positive rational integer. Bianchi groups are of major importance, as they form a representative set for the commensurability classes of all non-cocompact arithmetic lattices in three-dimensional hyperbolic space [17, 18]. Furthermore we investigate certain congruence subgroups of $\mathrm{SL}(2, \mathfrak{o}_K)$. The case $D = 1$ is discussed in detail in section VI, together with those cocompact arithmetic groups, which are of index two in the Coxeter groups $T_i$, $i = 1, 2, 4, 5, 7, 9$, generated by the reflections at the faces of corresponding Lannér's hyperbolic tetrahedra.

We shall focus our attention on arithmetic quaternion groups $\mathcal{O}^1$, as they form the species of groups with the strongest arithmetic structure. Anyway, from our detailed investigation of the length spectra of arithmetic three-orbifolds connected with such groups we easily obtain bounds for $\mathcal{N}^r(l)$ and $\langle g(l) \rangle$ for arbitrary arithmetic three-orbifolds using the characterization of arithmetic Kleinian groups by Maclachlan and Reid [17]. However, the relations between the length spectra associated with commensurable groups are in general not as simple as stated in [9]; we give a counter example in section VII.

## II   Closed geodesics in hyperbolic three-orbifolds

The natural extension of the half plane model $\mathfrak{H}_2$ of two-dimensional hyperbolic space $\mathbb{H}^2$ to three-dimensional hyperbolic space $\mathbb{H}^3$ is the *upper half space*

$$(7) \qquad \mathfrak{H}_3 = \{(x_1, x_2, x_3) \in \mathbb{R}^3 \mid x_3 > 0\},$$

equipped with the Riemannian metric

$$(8) \qquad ds^2 = \kappa \, \frac{dx_1^2 + dx_2^2 + dx_3^2}{x_3^2}.$$

The curvature is $-1/\kappa$ everywhere on $\mathfrak{H}_3$; we choose $\kappa = 1$. The hyperbolic distance $d(\boldsymbol{x}, \boldsymbol{x}')$ of two points $\boldsymbol{x}, \boldsymbol{x}' \in \mathfrak{H}_3$ is then given by

$$(9) \qquad \cosh d(\boldsymbol{x}, \boldsymbol{x}') = 1 + \frac{(x_1 - x_1')^2 + (x_2 - x_2')^2 + (x_3 - x_3')^2}{2 x_3 x_3'}.$$

Compared to other models of three-dimensional hyperbolic space $\mathbb{H}^3$ (see, e.g., [19]), the main advantage of this model is that the action of the group of isometries has a simple representation by fractional linear transformations: Set $\boldsymbol{x} = x_1 + x_2\,\mathrm{i} + x_3\,\mathrm{j}$, with the classical quaternions i and j defined by the relations $\mathrm{i}^2 = \mathrm{j}^2 = -1$ and $\mathrm{ij} + \mathrm{ji} = 0$, plus the property that i and j commute with every real number. 1, i, j and ij form a basis for the classical Hamilton quaternions $\mathbb{H}$. The inverse of a quaternion $q = q_1 + q_2\,\mathrm{i} + q_3\,\mathrm{j} + q_4\,\mathrm{ij}$ in $\mathbb{H}$ is given by $q^{-1} = |q|^{-2}(q_1 - q_2\,\mathrm{i} - q_3\,\mathrm{j} - q_4\,\mathrm{ij})$, where $|q|^2 = q_1^2 + q_2^2 + q_3^2 + q_4^2$. Then every orientation-preserving isometry $f$ of $\mathfrak{H}_3$ has a representation

$$(10) \qquad f(x) = (ax + b)(cx + d)^{-1}, \text{ where } \begin{pmatrix} a & b \\ c & d \end{pmatrix} \in \mathrm{SL}(2, \mathbb{C}),$$

while for the orientation-reversing case this matrix should be contained in the coset $\mathrm{SL}(2, \mathbb{C})\,\mathrm{j}$. The group of orientation-preserving isometries $\mathrm{Iso}^+\,\mathfrak{H}_3$ is thus isomorphic to $\mathrm{PSL}(2, \mathbb{C}) = \mathrm{SL}(2, \mathbb{C})/\{\pm 1\}$, the class of orientation-reversing isometries $\mathrm{Iso}^-\,\mathfrak{H}_3$ is isomorphic to $\mathrm{PSL}(2, \mathbb{C})\,\mathrm{j}$, cf. [20]. Therefore the action of $G := \mathrm{SL}(2, \mathbb{C}) \cup \mathrm{SL}(2, \mathbb{C})\,\mathrm{j}$ on $\mathfrak{H}_3$ can be defined as the action of the associated isometry. Elements of $G$, and hence of $\mathrm{Iso}\,\mathfrak{H}_3$, can be classified as follows:

$g \in G$ is called ... ,      if it is $\mathrm{SL}(2, \mathbb{C})$-conjugate to ...

*plane reflection* $\qquad \pm \begin{pmatrix} 1 & 0 \\ 0 & 1 \end{pmatrix} \mathrm{j}$

*elliptic* $\qquad \pm \begin{pmatrix} \mathrm{e}^{\mathrm{i}\phi/2} & 0 \\ 0 & \mathrm{e}^{-\mathrm{i}\phi/2} \end{pmatrix}, \qquad \phi \in (0, \pi]$

*inverse elliptic* $\qquad \pm \begin{pmatrix} 0 & \mathrm{i}\,\mathrm{e}^{\mathrm{i}\phi/2} \\ \mathrm{i}\,\mathrm{e}^{-\mathrm{i}\phi/2} & 0 \end{pmatrix} \mathrm{j}, \qquad \phi \in (0, \pi]$

*parabolic* $\qquad \pm \begin{pmatrix} 1 & 1 \\ 0 & 1 \end{pmatrix}$

*inverse parabolic* $\qquad \pm \begin{pmatrix} 1 & 1 \\ 0 & 1 \end{pmatrix} \mathrm{j}$

*hyperbolic* $\qquad \pm \begin{pmatrix} \mathrm{e}^{l/2 + \mathrm{i}\phi/2} & 0 \\ 0 & \mathrm{e}^{-l/2 - \mathrm{i}\phi/2} \end{pmatrix}, \qquad l > 0, \qquad \phi \in [0, 2\pi)$

*inverse hyperbolic* $\qquad \pm \begin{pmatrix} \mathrm{e}^{l/2} & 0 \\ 0 & \mathrm{e}^{-l/2} \end{pmatrix} \mathrm{j}, \qquad l > 0.$

We call $l$ the *real length*, $\phi$ the *phase*, and $\ell := l + \mathrm{i}\phi$ the *complex length* of the transformation. Usually the real length is simply denoted as *the* length of the transformation, but as in this work also complex lengths will play an important role, we will keep the longer notation. Hyperbolic elements are called *loxodromic*, if $\phi \neq 0$.

To construct a hyperbolic manifold or – to be more general – a hyperbolic orbifold, let us take a discrete subgroup $\Gamma$ of $G$, and identify all points of $\mathfrak{H}_3$, which can be transformed into each other by an element of $\Gamma$. Such points are called $\Gamma$-*equivalent* and we put them into an equivalence class $\Gamma(\boldsymbol{x}) = \{\gamma(\boldsymbol{x}) \mid \gamma \in \Gamma\}$ with $\boldsymbol{x} \in \mathfrak{H}_3$. The set of those classes is the hyperbolic three-orbifold $\Gamma\backslash\mathfrak{H}_3 = \{\Gamma(\boldsymbol{x}) \mid \boldsymbol{x} \in \mathfrak{H}_3\}$. $\Gamma\backslash\mathfrak{H}_3$ is a manifold, if $\Gamma$ contains no elements of finite order other than $\pm 1$. We will call $\Gamma$ a hyperbolic *lattice*, if the volume of $\Gamma\backslash\mathfrak{H}_3$ is finite. If $\Gamma\backslash\mathfrak{H}_3$ is compact, $\Gamma$ will be called *cocompact* – which is the case, if and only if $\Gamma$ is a lattice containing no parabolic transformations. Discrete subgroups of $\mathrm{SL}(2,\mathbb{C})$ are denoted as *Kleinian* groups.

One way of visualizing a hyperbolic three-orbifold is to look at the fundamental cell of the corresponding lattice, where boundary points are identified by certain lattice transformations. For illustration, see section VI.

Every primitive closed geodesic $\mathcal{S}_0$ in $\Gamma\backslash\mathfrak{H}_3$ lifts to geodesics in $\mathfrak{H}_3$. Consider one of those lifts and the group $S$ of elements in $\Gamma$ leaving it invariant. Let $l_0$ denote the smallest length of all hyperbolic and inverse hyperbolic transformations in $S$. Then $l_0$ corresponds to the length of $\mathcal{S}_0$. The lengths of all other hyperbolic transformations in $S$ are of the form $l = k\, l_0$, $k \in \mathbb{N}$, (this follows from the discreteness of $S$) corresponding to lengths of closed geodesics $\mathcal{S}$ being multiple traversals of $\mathcal{S}_0$.

We will now give four simple, but very useful relations between the lengths and the traces of elements $\gamma$ in $\mathrm{SL}(2,\mathbb{C})$:

(11) $\qquad \mathrm{tr}\,\gamma = \pm 2\cosh(l/2)\cos(\phi/2) \pm 2\,\mathrm{i}\sinh(l/2)\sin(\phi/2),$

$$\text{(12)} \qquad \frac{\operatorname{Re}^2 \operatorname{tr} \gamma}{4\cosh^2(l/2)} + \frac{\operatorname{Im}^2 \operatorname{tr} \gamma}{4\sinh^2(l/2)} = 1,$$

$$\text{(13)} \qquad \frac{\operatorname{Re}^2 \operatorname{tr} \gamma}{4\cos^2(\phi/2)} - \frac{\operatorname{Im}^2 \operatorname{tr} \gamma}{4\sin^2(\phi/2)} = 1,$$

and

$$\text{(14)} \qquad x_\pm = \frac{1}{2}\left(|\operatorname{tr}^2 \gamma| \pm |\operatorname{tr}^2 \gamma - 4|\right),$$

where $x_+ := 2\cosh l$ and $x_- := 2\cos\phi$. These relations can be used for elements in $\mathrm{SL}(2,\mathbb{C})\,\mathrm{j}$, if one replaces $\gamma$ by $\gamma^2$, $l$ by $2l$, and $\phi$ by $2\phi$.

## III  Arithmetic hyperbolic three-orbifolds

Before giving a definition of arithmetic hyperbolic three-orbifolds, it is necessary to introduce the terms *quaternion algebra* and *order in a quaternion algebra*, cf. [21].

Let $K$ be a field, and $A$ an algebra over $K$ generated by the elements $1, \omega, \Omega, \omega\Omega$, with the relations

$$\text{(15)} \qquad \omega^2 = a, \quad \Omega^2 = b, \quad \omega\Omega + \Omega\omega = 0,$$

where $a, b \in K - \{0\}$. Then $A$ is called *quaternion algebra over $K$*. Every element $\alpha \in A$ has a unique representation

$$\alpha = x_0 + x_1\omega + x_2\Omega + x_3\omega\Omega, \qquad x_j \in K.$$

Then

$$\alpha^\dagger = x_0 - x_1\omega - x_2\Omega - x_3\omega\Omega$$

denotes the conjugate of $\alpha$. The reduced trace and norm are defined by

$$\text{(16)} \qquad \operatorname{tr}_A \alpha := \alpha + \alpha^\dagger = 2x_0$$

and

$$\text{n}_A\,\alpha := \alpha\alpha^\dagger = x_0^2 - x_1^2 a - x_2^2 b + x_3^2 ab. \tag{17}$$

The field extension $K(\omega)$ is isomorphic to $K(\sqrt{a})$, as $\omega$ commutes with all elements of $K$. The quaternion algebra $A$ is therefore isomorphic to the matrix algebra

$$\left\{ \begin{pmatrix} x_0 + x_1\sqrt{a} & x_2 + x_3\sqrt{a} \\ b\,(x_2 - x_3\sqrt{a}) & x_0 - x_1\sqrt{a} \end{pmatrix} \;\bigg|\; x_0,\ldots,x_3 \in K \right\}.$$

An *order* $\mathcal{O}$ in the quaternion algebra $A$ is a ring of elements of $A$ with

(i) the ring of integers $\mathfrak{o}_K$ in $K$ is contained in $\mathcal{O}$,

(ii) the reduced trace and norm of every element in $\mathcal{O}$ are integers in $K$,

(iii) $\mathcal{O}$ is generated by four elements that are linearly independent over $K$.

An order is called *maximal*, if it is not properly contained in any other order. It is easy to show that $\operatorname{tr}\mathcal{O} = \{\operatorname{tr}_A \alpha \mid \alpha \in \mathcal{O}\}$ is an ideal in $\mathfrak{o}_K$: Let $\alpha \in \mathcal{O}$. Then $r\alpha \in \mathcal{O}$ and $r\alpha^\dagger \in \mathcal{O}^\dagger$ for all $r \in \mathfrak{o}_K$, because $\mathcal{O}$ and $\mathcal{O}^\dagger = \{\alpha^\dagger \mid \alpha \in \mathcal{O}\}$ are rings. It follows that $r\operatorname{tr}_A \alpha \in \operatorname{tr}\mathcal{O}$ for all $r \in \mathfrak{o}_K$. Together with $\operatorname{tr}_A \alpha$ and $\operatorname{tr}_A \beta$ also $(\operatorname{tr}_A \alpha - \operatorname{tr}_A \beta)$ is contained in $\operatorname{tr}\mathcal{O}$.

Now a hyperbolic orbifold is called *arithmetic*, if the corresponding lattice $\Gamma$ is arithmetic. *Arithmetic groups* in $G$ are defined as follows [17]: Let $K$ be a number field of degree $d$ over $\mathbb{Q}$ with exactly one complex place. Let $\phi_j : K \hookrightarrow \mathbb{C}$, $j = 1,\ldots,d$ denote the embeddings of $K$ into the field of complex numbers, such that $\phi_1 =$ identity (id), $\phi_2 =$ complex conjugation (cc), $\phi_j(K) \subset \mathbb{R}$ for $j = 3,\ldots,d$. Let furthermore $A$ be a quaternion algebra over $K$. It splits at the complex place, i.e.,

$$A \otimes_{\phi_1(K)} \mathbb{C} \simeq \mathrm{M}(2,\mathbb{C}). \tag{18}$$

We demand that $A$ is ramified at the real places,

$$A \otimes_{\phi_j(K)} \mathbb{R} \simeq \mathbb{H}, \qquad \text{for } j = 3,\ldots,d\,. \tag{19}$$

It then follows that

(20) $$A \otimes_{\mathbb{Q}} \mathbb{R} \simeq \mathrm{M}(2, \mathbb{C}) \oplus \mathbb{H} \oplus \ldots \oplus \mathbb{H}.$$

Let $\rho_1$ be the projection onto the first summand restricted to $A$, $\rho_3$ the projection onto the second one, etc.,

(21) $$\rho_1 : A \to \mathrm{M}(2, \mathbb{C}),$$

(22) $$\rho_j : A \to \mathbb{H}, \qquad \text{for } j = 3, \ldots, d .$$

If $\mathcal{O}$ is an order in $A$ and $\mathcal{O}^1$ the group of units of $\mathcal{O}$ with norm one, then $\rho_1(\mathcal{O}^1)$ is a lattice in $\mathrm{SL}(2, \mathbb{C})$, which we will denote as an *arithmetic quaternion group*. Every group commensurable with $\rho_1(\mathcal{O}^1)$ is called *arithmetic*, i.e., if $\Gamma \cap \rho_1(\mathcal{O}^1)$ is of finite index in both $\Gamma$ and $\rho_1(\mathcal{O}^1)$. A subgroup of finite index in $\rho_1(\mathcal{O}^1)$ is said to be *derived from a quaternion algebra*. In the following we will briefly write $\mathcal{O}^1$ instead of $\rho_1(\mathcal{O}^1)$.

The trace of an element $\gamma \in \Gamma$ is given by

(23) $$\operatorname{tr} \gamma = \operatorname{tr}_{\mathrm{M}(2,\mathbb{C})} \rho_1(\alpha) = \operatorname{tr}_A \alpha,$$

and the conjugates of the trace by

(24) $$\phi_j(\operatorname{tr} \gamma) = \operatorname{tr}_{\mathbb{H}} \rho_j(\alpha).$$

For the determinant we obtain

(25) $$\det \gamma = \mathrm{n}_{\mathrm{M}(2,\mathbb{C})} \rho_1(\alpha) = \mathrm{n}_A \alpha,$$

(26) $$\phi_j(\det \gamma) = \mathrm{n}_{\mathbb{H}} \rho_j(\alpha).$$

The trace field $\mathbb{Q}(\operatorname{tr} \Gamma)$ is naturally contained in $K$. Moreover, Maclachlan and Reid [17] have shown that $K$ even coincides with $\mathbb{Q}(\operatorname{tr} \Gamma)$. It follows from its definition that an arithmetic group is always of finite covolume, i.e., a lattice. It is cocompact if and only if $A$ is a division algebra.

As was done by Takeuchi for Fuchsian groups [22], it is possible to characterize arithmetic Kleinian groups by looking at their traces:

**Theorem 1** [17] *$\Gamma$ is an arithmetic Kleinian group, if and only if $\Gamma^{(2)} = \langle \gamma^2 | \gamma \in \Gamma \rangle$ (the group generated by the squares of elements of $\Gamma$) is derived from a quaternion algebra.*

It is clear that a lattice in $G$ is arithmetic, if and only if its subgroup of all orientation-preserving elements is an arithmetic Kleinian group.

**Theorem 2** [17] *Let $\Gamma$ be a Kleinian group of finite covolume. Then $\Gamma$ is derived from a quaternion algebra, if and only if*

(i) *$K = \mathbb{Q}(\operatorname{tr} \Gamma)$ is an algebraic number field, such that $K \not\subset \mathbb{R}$ and $\operatorname{tr} \Gamma \subset \mathfrak{o}_K$,*

(ii) *for every embedding $\phi : K \hookrightarrow \mathbb{C}$ such that $\phi \notin \{\operatorname{id}, \operatorname{cc}\}$, $\phi(\operatorname{tr} \Gamma)$ is bounded in $\mathbb{C}$.*

REMARKS: We can put condition (ii) into a more concrete form: $K$ coincides with the field $K'$, over which $A$ is defined, because there cannot be a proper complex subfield of $K'$ ($K'$ has just one complex place, see also lemma 2). Hence all embeddings $\phi \notin \{\operatorname{id}, \operatorname{cc}\}$ are real. It is easy to see that $\phi(\operatorname{tr} \Gamma)$ is contained in the interval $[-2, 2]$, following from the fact that $A \otimes_{\phi(K)} \mathbb{R}$ is isomorphic to the Hamilton quaternions $\mathbb{H}$.

Next consider an arbitrary arithmetic Kleinian group $\Gamma$. It follows from the relation

$$(27) \qquad (\operatorname{tr} \gamma)^2 = \operatorname{tr} \gamma^2 + 2,$$

that every $\operatorname{tr} \gamma \in \operatorname{tr} \Gamma$ is the square-root of an integer in $\mathbb{Q}(\operatorname{tr} \Gamma^{(2)})$. As $\Gamma^{(2)}$ is derived from a quaternion algebra, for every embedding $\phi$ of $\mathbb{Q}(\operatorname{tr} \Gamma^{(2)})$ into $\mathbb{C}$ such that $\phi \notin \{\operatorname{id}, \operatorname{cc}\}$, $\phi(\{\operatorname{tr} \gamma^2 \mid \gamma \in \Gamma\})$ is contained in the interval $[-2, 2]$, and equivalently $\phi(\operatorname{tr}^2 \Gamma)$ is contained in the interval $[0, 4]$, where $\operatorname{tr}^2 \Gamma$ denotes the set of squared traces $(\operatorname{tr} \gamma)^2$ of $\Gamma$.

In the case of a general arithmetic lattice $\Gamma$ in $G$ denote by $\Gamma^+$ the subgroup of all orientation-preserving elements. Then $\gamma^2 \in \Gamma^+$ for all $\gamma \in \Gamma$, so for every embedding $\phi$ of $\mathbb{Q}(\operatorname{tr}\Gamma^{+(2)})$ into $\mathbb{C}$ such that $\phi \notin \{\operatorname{id}, \operatorname{cc}\}$, $\phi(\{\operatorname{tr}\gamma^4 \mid \gamma \in \Gamma\})$ is contained in the interval $[-2, 2]$.

## IV  Number of distinct complex lengths

By relation (11) to each trace $\operatorname{tr}\gamma$ of an element $\gamma$ in a Kleinian group $\Gamma$ we can associate a complex length $\ell_\gamma = l_\gamma + \mathrm{i}\phi_\gamma$. Vice versa, to each complex length $\ell_\gamma$ there correspond exactly two traces $\pm\operatorname{tr}\gamma$, if we assume that $-1 \in \Gamma$. This is true, e.g., for all arithmetic quaternion groups. If $-1$ is not contained in $\Gamma$, simply replace $\Gamma$ by $\Gamma \cup \Gamma(-1)$, which has no effect on the length spectrum. Hence the number of distinct complex lengths of closed geodesics in $\Gamma\backslash\mathfrak{H}_3$,

$$\tag{28} \mathcal{N}^c(l) := \#\{\ell_\gamma = l_\gamma + \mathrm{i}\phi_\gamma \mid 0 < l_\gamma \leq l,\ \gamma \in \Gamma\},$$

is given by

$$\tag{29} \mathcal{N}^c(l) = \frac{1}{2}\#\left\{\operatorname{tr}\gamma \,\bigg|\, \frac{\operatorname{Re}^2 \operatorname{tr}\gamma}{4\cosh^2(l/2)} + \frac{\operatorname{Im}^2 \operatorname{tr}\gamma}{4\sinh^2(l/2)} \leq 1,\ \gamma \in \Gamma\ \text{hyperbolic}\right\}.$$

Let us introduce the function

$$\tag{30} \mathcal{A}(x) := \#\left\{\operatorname{tr}\gamma \,\bigg|\, \frac{\operatorname{Re}^2 \operatorname{tr}\gamma}{x+2} + \frac{\operatorname{Im}^2 \operatorname{tr}\gamma}{x-2} \leq 1,\ \gamma \in \Gamma\right\}.$$

$\mathcal{A}(x)$ additionally counts the distinct traces in the interval $[-2, 2]$, i.e., traces of trivial, elliptic and parabolic elements in $\Gamma$. Let $N$ be the number of these traces (which is finite, since there are only finitely many elliptic conjugacy classes), then

$$\tag{31} \mathcal{N}^c(l) = \frac{1}{2}\left[\mathcal{A}(2\cosh l) - N\right].$$

First let us concentrate on the length spectrum of an arithmetic quaternion group $\mathcal{O}^1$: For $j = 3, \ldots, d$, all conjugates $\phi_j(\operatorname{tr}\gamma)$ of traces are contained in the interval $[-2, 2]$, see above. Let $\mathfrak{a} := \operatorname{tr}\mathcal{O}$ and

$$\tag{32} \mathfrak{a}_\square := \{s \in \mathfrak{a} \mid\ -2 \leq \phi_j(s) \leq 2,\ j = 3, \ldots, d\}.$$

We divide $\mathcal{A}(x)$ into two parts:

$$\mathcal{A}(x) = \mathcal{E}(x) - \mathcal{G}(x), \tag{33}$$

where

$$\mathcal{E}(x) := \#\left\{ s \in \mathfrak{a}_\square \ \Big|\ \frac{\operatorname{Re}^2 s}{x+2} + \frac{\operatorname{Im}^2 s}{x-2} \leq 1 \right\} \tag{34}$$

and

$$\mathcal{G}(x) := \#\left\{ s \in \mathfrak{g} \ \Big|\ \frac{\operatorname{Re}^2 s}{x+2} + \frac{\operatorname{Im}^2 s}{x-2} \leq 1 \right\}, \tag{35}$$

where $\mathfrak{g}$ is the set of all elements $s \in \mathfrak{a}_\square$, which do not represent a trace of an element of $\mathcal{O}^1$, i.e.,

$$\mathfrak{g} = \mathfrak{a}_\square - \operatorname{tr} \mathcal{O}^1. \tag{36}$$

The elements of $\mathfrak{g}$ will be called *gaps in* $\operatorname{tr} \mathcal{O}^1$, the corresponding lengths *gaps in the length spectrum*, as their number is conjectured to be small compared to $\mathcal{E}(x)$ and $\mathcal{N}^c(l)$, respectively (see the conjecture below). It is easy to see that in the case of lattices $\mathcal{O}^1$ isomorphic to a Bianchi group $\Gamma = \operatorname{SL}(2, \mathfrak{o}_K)$, where $\mathfrak{o}_K$ is the ring of integers of the quadratic number field $K = \mathbb{Q}(\sqrt{-D})$, $D$ a square-free positive rational integer [17, 18], no gaps can occur, since $\operatorname{tr} \Gamma = \mathfrak{o}_K$ (compare lemma 1), hence $\mathcal{G}(x) = 0$.

Let us return to the general case, where $\mathcal{O}^1$ is an arbitrary arithmetic quaternion group in $\operatorname{SL}(2, \mathbb{C})$. We will compute the leading order of $\mathcal{E}(x)$ as $x \to \infty$: Every number $s \in K$ can be represented geometrically as a vector $\boldsymbol{s}$ in $\mathbb{C} \times \mathbb{R}^{d-2}$, see [23]:

$$\boldsymbol{s} \equiv \begin{pmatrix} s_1 \\ s_3 \\ \vdots \\ s_d \end{pmatrix} := \begin{pmatrix} \phi_1(s) \\ \phi_3(s) \\ \vdots \\ \phi_d(s) \end{pmatrix}.$$

Thus the components of $\boldsymbol{s}$ are the conjugates of $s$. We call $\boldsymbol{s}$ the geometric image of $s$.

If we represent a $\mathbb{Z}$-module geometrically, we will obtain a Euclidean lattice in $\mathbb{C} \times \mathbb{R}^{d-2}$, whose fundamental cell is spanned by the geometric images of the basis elements of the module.

As stated above, $\mathfrak{a} = \operatorname{tr} \mathcal{O}$ is an ideal in $\mathfrak{o}_K$. Every ideal is an $\mathfrak{o}_K$-module and hence a *complete* $\mathbb{Z}$-module, i.e., there are $d$ linearly independent basis elements. We denote the geometric image of $\mathfrak{a}$ by $\mathfrak{M}$ and its fundamental cell by $\mathcal{F}_\mathfrak{M}$.

Now $\mathcal{E}(x)$ corresponds to the number of lattice points, contained in the elliptic cylinder

$$(37) \quad E(x) := \left\{ s \in \mathbb{C} \times \mathbb{R}^{d-2} \,\bigg|\, \frac{\operatorname{Re}^2 s_1}{x+2} + \frac{\operatorname{Im}^2 s_1}{x-2} \leq 1,\ -2 \leq s_j \leq 2,\ j = 3, \ldots, d \right\}.$$

Let $\mathcal{Z}(x)$ be the number of lattice points in the cylinder

$$(38) \quad Z(x) := \left\{ s \in \mathbb{C} \times \mathbb{R}^{d-2} \,\big|\, |s_1|^2 \leq x,\ |s_j| \leq 2,\ j = 3, \ldots, d \right\},$$

then

$$(39) \quad \mathcal{Z}(x-2) \leq \mathcal{E}(x) \leq \mathcal{Z}(x+2).$$

The leading order of $\mathcal{Z}(x)$ is given by a classical result, see, e.g., theorem 1 of chapter V in [24],

$$(40) \quad \mathcal{Z}(x) = \frac{\operatorname{Vol} Z(x)}{\operatorname{Vol} \mathcal{F}_\mathfrak{M}} + O(x^{1-1/d}), \qquad x \to \infty.$$

with $\operatorname{Vol} Z(x) = 4^{d-2}\pi x$ and $\operatorname{Vol} \mathcal{F}_\mathfrak{M} = 2^{-1}|D_\mathfrak{a}|^{1/2}$, and the discriminant $D_\mathfrak{a}$ of $\mathfrak{a}$, hence

$$(41) \quad \mathcal{Z}(x) = \frac{2^{2d-3}\pi}{|D_\mathfrak{a}|^{1/2}} x + O(x^{1-1/d}), \qquad x \to \infty.$$

Because of (39) we have

$$(42) \quad \mathcal{Z}(x) = \mathcal{Z}(x \pm 2) + O(x^{1-1/d}) = \mathcal{E}(x) + O(x^{1-1/d}), \qquad x \to \infty,$$

which gives us the following theorem:

**Theorem 3** *Let $\mathcal{O}^1$ be an arithmetic quaternion group in $\mathrm{SL}(2,\mathbb{C})$. Then the number of distinct complex lengths of closed geodesics is given by*

$$\mathcal{N}^c(l) = \frac{2^{2d-4}\pi}{|D_a|^{1/2}}\, \mathrm{e}^l - \frac{1}{2}\mathcal{G}(\mathrm{e}^l) + O(\mathrm{e}^{l(1-1/d)}), \qquad l \to \infty. \tag{43}$$

It should be noted that in the non-compact case ($d = 2$) $\mathscr{Z}(x)$ is the number of lattice points inside a circle of radius $x^{1/2}$. The estimate of the remainder function has a long history in number theory, and there are better results than our trivial $O(x^{1/2})$, going originally back to Gauß. To the author's knowledge the best estimate for a quadratic lattice (corresponding to $\mathbb{Z}[i]$) is by Huxley [25], where the remainder function is of order $O(x^{23/73+\epsilon})$, for any number $\epsilon > 0$. However, the search for a sharp estimate – known as "the classical circle problem" – is still going on.

There has not been a lot of progress in finding asymptotic bounds for the number of gaps $\mathcal{G}(x)$, even in the more simple two-dimensional case. (An explicit list of gaps in the case of the regular octagon group may be found in [11].) Anyway, the following conjecture seems to be quite natural.

**Conjecture 1** *Let $\mathcal{O}^1$ be an arithmetic quaternion group in $\mathrm{SL}(2,\mathbb{C})$. Then the number of gaps in the complex length spectrum up to length $l = \log x$ is given by*

$$\mathcal{G}(x) = \kappa\, x + o(x), \qquad x \to \infty, \tag{44}$$

*where $\kappa \geq 0$ is a constant depending only on $\mathcal{O}^1$, and small compared to*

$$\frac{2^{2d-3}\pi}{|D_a|^{1/2}}\,.$$

One might suspect that even $\kappa = 0$ for all $\mathcal{O}^1$.

For a group $\Gamma$, which is derived from a quaternion algebra and therefore is a subgroup of finite index in an arithmetic quaternion group $\mathcal{O}^1$, one trivially has $\mathcal{N}^c(l;\Gamma) \leq \mathcal{N}^c(l;\mathcal{O}^1)$. A more precise relation is not known in general. However, if one considers congruence subgroups of a Bianchi group, a lot more can be said. Let us define the principal congruence subgroup

$$(45)\quad \Gamma(\mathfrak{n}) = \left\{ \begin{pmatrix} a & b \\ c & d \end{pmatrix} \in \mathrm{SL}(2, \mathfrak{o}_K) \;\bigg|\; \begin{pmatrix} a & b \\ c & d \end{pmatrix} \equiv \begin{pmatrix} 1 & 0 \\ 0 & 1 \end{pmatrix} \mod \mathfrak{n} \right\},$$

where $\mathfrak{n}$ is an ideal in $\mathfrak{o}_K$. For simplicity we will assume that $\mathfrak{n}$ is a principal ideal, i.e., $\mathfrak{n} = (N)$, $N \in \mathfrak{o}_K$. Then we are able to prove

**Lemma 1** *An element $s \in \mathfrak{o}_K$ is the trace of an element in $\Gamma(\mathfrak{n})$, $\mathfrak{n} = (N)$ a principal ideal, if and only if $s \equiv 2 \mod N^2$.*

PROOF: As $bc \equiv 0 \mod N^2$, clearly $ad \equiv 1 \mod N^2$. Now write $a$ and $d$ in the form $a = Nj + 1$ and $d = Nk + 1$, where $j, k \in \mathfrak{o}_K$. It follows that $N(j+k) \equiv 0 \mod N^2$. Then $\mathrm{tr}\,\gamma = a + d = N(j+k) + 2 \equiv 2 \mod N^2$ is a necessary condition. Now consider the set

$$M = \left\{ \begin{pmatrix} N^2 j + 1 & Nj \\ N & 1 \end{pmatrix} \;\bigg|\; j \in \mathfrak{o}_K \right\},$$

which is contained in $\Gamma(\mathfrak{n})$ and whose set of traces $\mathrm{tr}\,M$ contains every desired element. $\square$

Similar results can be obtained for other congruence subgroups, e.g.,

$$(46)\quad \Gamma_1(\mathfrak{n}) = \left\{ \begin{pmatrix} a & b \\ c & d \end{pmatrix} \in \mathrm{SL}(2, \mathfrak{o}_K) \;\bigg|\; a \equiv d \equiv 1,\; c \equiv 0 \mod \mathfrak{n} \right\},$$

where each $s \in \mathfrak{o}_K$ with $s \equiv 2 \mod N$ is a trace of an element of $\Gamma_1(\mathfrak{n})$, if $\mathfrak{n} = (N)$. Theorem 3 can therefore as well be applied for such congruence groups, e.g., by setting $\mathfrak{a} = (N^2)$ and $\mathcal{G}(x) = 0$ in the case of $\Gamma(\mathfrak{n})$, if $N$ divides 2, i.e., $2/N \in \mathfrak{o}_K$. If $N$ does not divide 2, $s \equiv 2 \mod N^2$ implies $-s \not\equiv 2 \mod N^2$, hence in this case there is an additional factor 2:

$$(47)\quad \mathcal{N}^c(l; \Gamma(\mathfrak{n})) = \begin{cases} \dfrac{1}{|N|^4} \mathcal{N}^c(l; \mathrm{SL}(2, \mathfrak{o}_K)) + O(\mathrm{e}^{l/2}), & \text{if } N \text{ divides } 2 \\[2mm] \dfrac{2}{|N|^4} \mathcal{N}^c(l; \mathrm{SL}(2, \mathfrak{o}_K)) + O(\mathrm{e}^{l/2}), & \text{otherwise.} \end{cases}$$

Unfortunately the methods applied above cannot produce analogous results in the general setting. In any case, from our remarks on Maclachlan and Reid's theorems 1 and 2 in section III it follows that the number of distinct complex lengths of every arithmetic hyperbolic three-orbifold is of the order

$$\mathcal{N}^c(l) = O(\mathrm{e}^l), \qquad l \to \infty. \tag{48}$$

## V  Number of distinct real lengths. Multiplicities

As we have seen, all traces $\mathrm{tr}\,\gamma$ belonging to the same real length $l$ are located on the ellipse

$$e(x) = \left\{ s \in \mathbb{C} \,\middle|\, \frac{\mathrm{Re}^2\, s}{x+2} + \frac{\mathrm{Im}^2\, s}{x-2} = 1 \right\}, \tag{49}$$

with $x = 2\cosh l$. Hence the number of distinct real lengths

$$\mathcal{N}^r(l) = \#\{l_\gamma \in (0, l] \mid \gamma \in \Gamma\} \tag{50}$$

corresponds to the number of ellipses $e(x)$, each of which contains at least one trace $\mathrm{tr}\,\gamma$. Defining

$$\mathcal{B}(x) := \#\{e(y) \mid \mathfrak{a}_\Box \cap e(y) \neq \emptyset,\ y \leq x\}, \tag{51}$$

we immediately see that

$$\mathcal{N}^r(l) = \mathcal{B}(2\cosh l) - \#\text{ gaps in the real length spectrum up to } l. \tag{52}$$

In the following we will restrict ourselves to those $\mathcal{O}^1$, whose sets of traces $\mathrm{tr}\,\mathcal{O}^1$ are invariant under complex conjugation. By that the algebraic set-up is simplified considerably, as can be drawn from the following lemmata. For example, the subgroup of orientation-preserving elements in an arithmetic lattice, generated by the reflections at the faces of a polyhedron, satisfies this condition. We will also give weaker results for general arithmetic lattices in $G$ at the end of this section.

**Lemma 2** *Let $K$ be a number field with exactly one complex place. If $K$ is invariant under complex conjugation, then there exists a real subfield $L$, such that the degree of $K$ over $L$ is two.*

PROOF (compare [17]): *Step 1.* We show that each proper subfield $M$ of $K$ is real.

Suppose, $M \neq K$ is a complex subfield of $K$, $[K : M] = \tilde{d}$. Then $M$ has one complex place. We have denoted the embeddings of $K$ into $\mathbb{C}$ by $\phi_1 = \mathrm{id}$, $\phi_2 = \mathrm{cc}$, $\phi_j(K) \subset \mathbb{R}$ for $j = 3, \ldots, d$. Let the embeddings of $M$ into $\mathbb{C}$ be $\hat{\phi}_1 = \mathrm{id}$, $\hat{\phi}_2 = \mathrm{cc}$, $\hat{\phi}_j(M) \subset \mathbb{R}$ for $j = 3, \ldots, d/\tilde{d}$.

Of course $\phi_1|_M = \hat{\phi}_1$, $\phi_2|_M = \hat{\phi}_2$. There has to be at least one further $\phi_j$, such that $\phi_j|_M = \hat{\phi}_1$. But as $\phi_j(K) \subset \mathbb{R}$ for $j = 3, \ldots, d$, it follows that $\phi_j(M) = \hat{\phi}_1(M) = M \subset \mathbb{R}$, contradiction!

*Step 2.* Let $K = \mathbb{Q}(a)$, $a \in \mathbb{C}$. As $K$ is invariant under complex conjugation, we have $\overline{a} \in K$ and $\mathrm{i}\, b := a - \overline{a} \in K$. $K$ is the only non-real subfield of $K$, thus $K = \mathbb{Q}(\mathrm{i}\, b)$.

Let $L = K \cap \mathbb{R}$ be the maximal real subfield of $K$, then $D := -(\mathrm{i}\, b)^2 > 0$ is in $L$, hence $K = L(\sqrt{-D})$ is of degree two over $L$. $\square$

We conclude that every number $s \in K$ has a representation $s = t + u\sqrt{-D}$, $t, u \in L$, $D > 0$ fixed in $L$.

We will now look at lattice points $s \in \mathrm{tr}\,\mathcal{O}^1$ on ellipses $e(x)$, for which $x \notin L$ (lemma 3), and at points on ellipses, for which $x \in L$ (lemma 4).

**Lemma 3** *Let $K$, $L$ be as above. Let $e(x)$, $x \notin L$, be an ellipse and $s \in \mathrm{tr}\,\mathcal{O}^1$ a point on it. Then there are exactly three additional points of $\mathrm{tr}\,\mathcal{O}^1$ on $e(x)$.*

PROOF: Let $s = t + u\sqrt{-D}$, $t, u \in L$ be on the ellipse. Thus
$$\frac{t^2}{x+2} + D\,\frac{u^2}{x-2} = 1.$$

Let $s' = t' + u'\sqrt{-D} \in \mathfrak{a}_\square$ be another point on $e(x)$, then

$$t^2 - t'^2 + D\,\frac{x+2}{x-2}\,(u^2 - u'^2) = 0.$$

As $x \notin L$, the same is true for

$$D\,\frac{x+2}{x-2} = D\left(1 + \frac{4}{x-2}\right) \notin L.$$

The differences $t^2 - t'^2$ and $u^2 - u'^2$ hence have to vanish, only the numbers $s$, $\overline{s}$, $-s$ and $-\overline{s}$ are on the ellipse $e(x)$. $s$, $\overline{s}$, $-s$ and $-\overline{s}$ are pairwise distinct, if $\operatorname{Re} s \neq 0$ and $\operatorname{Im} s \neq 0$. Indeed, this is the case – otherwise $\operatorname{Im}^2 s = x - 2$ or $\operatorname{Re}^2 s = x + 2$, and therefore $x \in L$, which contradicts our assumption. $\square$

**Lemma 4** *The counting function of all points $s \in \operatorname{tr} \mathcal{O}^1$ that are located on the set of ellipses $\{e(y) \mid 2 < y \leq x,\ y \in L\}$ is of order $O(x^{1/2})$, $x \to \infty$.*

PROOF: Let $s = t + u\sqrt{-D} \in \mathfrak{a}_\square$ be on $e(y)$. The equation, which describes the ellipse, is quadratic in $y$:

$$y^2 - (t^2 + D\,u^2)y + 2(t^2 - D\,u^2) - 4 = 0,$$

or equivalently,

(53) $$y^2 - |s^2|\,y + 2\operatorname{Re} s^2 - 4 = 0.$$

$s$ and $\overline{s}$ are contained in the ring of integers $\mathfrak{o}_K$, hence $|s^2| = s\overline{s}$ and $2\operatorname{Re} s^2 = s^2 + \overline{s}^2$ are also contained in $\mathfrak{o}_K$ and therefore are real algebraic integers of $L$, $|s^2| \in \mathfrak{o}_L$ and $2\operatorname{Re} s^2 \in \mathfrak{o}_L$.

$y$ is a zero of a normalized polynomial with integer coefficients contained in $\mathfrak{o}_L$, hence $y$ is an algebraic integer. As $y \in L$, we have $y \in \mathfrak{o}_L$.

For $s \neq \pm 2$ the quadratic equation (53) has two distinct solutions, compare (11),

(54) $$y_\pm = \frac{1}{2}\left[|s^2| \pm \sqrt{|s^2|^2 - 8\operatorname{Re} s^2 + 16}\right] = \frac{1}{2}\left(|s^2| \pm |s^2 - 4|\right).$$

Clearly $-2 \leq y_- \leq 2$ and $y_+ > 2$. We have $y_- = \pm 2$ if and only if $\operatorname{Im} s = 0$ resp. $\operatorname{Re} s = 0$. For $y_- \neq \pm 2$, $e(y_-)$ is a hyperbola, and $e(y_+)$ is an ellipse.

Conclusion: The number of points $s \in \mathfrak{a}_\square$ that are located on the set of ellipses $e(y)$, with $2 < y \leq x$ and $y \in L$, corresponds to the number of points $s \in \mathfrak{a}_\square$ that are located on the real axis, the imaginary axis, and the hyperbolas $e(y)$, $-2 < y < 2$, $y \in L$, and for which

$$\text{(55)} \qquad \frac{\operatorname{Re}^2 s}{x+2} + \frac{\operatorname{Im}^2 s}{x-2} \leq 1.$$

To give an estimate for that number, consider the geometric image of $y \in L$. Let $\hat{\phi}_1 = \operatorname{id}$, $\hat{\phi}_k$, $k = 2, \ldots, d/2$ be the embeddings of $L$ in $\mathbb{R}$ and $y_k = \hat{\phi}_k(y)$ the components of a vector $\boldsymbol{y} \in \mathbb{R}^{d/2}$. ($d$ is even, since $K$ is of degree two over $L$.)

The identity embedding of $L$ lifts to the identity $\phi_1$ and the complex conjugation $\phi_2$ of $K$, every other embedding of $L$ lifts to two corresponding embeddings $\phi_j$ of $K$. Now we have

$$-2 \leq \phi_j(s) \leq 2, \ j = 3, \ldots, d,$$

because $s \in \mathfrak{a}_\square$. The same is true for $\overline{s}$, thus $\phi_j(|s^2|) = \phi_j(s\overline{s})$ and $\phi_j(2 \operatorname{Re} s^2) = \phi_j(s^2 + \overline{s}^2)$ are bounded for $j = 3, \ldots, d$. In that case the solutions $\phi_j(y)$ of

$$\phi_j(y)^2 - \phi_j(|s^2|)\phi_j(y) + 2\phi_j(2 \operatorname{Re} s^2) - 4 = 0$$

are bounded, as they depend continuously on the coefficients. $y$ is contained in the ring of integers $\mathfrak{o}_L$, the vectors $\boldsymbol{y}$ describe lattice points in $\mathbb{R}^{d/2}$, whose components $y_2, \ldots, y_{d/2}$ are bounded, as we have seen. $y$ determines a hyperbola, if it is contained in the interval $(-2, 2)$. Therefore in this case the first component $y_1 = y$ is bounded, too. We conclude that the number of hyperbolas, which contain points $s \in \mathfrak{a}_\square$, is finite.

Let us pick one fixed hyperbola, and determine the number of points $s \in \mathfrak{a}_\square$ on it, which satisfy (55). A trivial estimate yields $O(x^{1/2})$, $x \to \infty$. For the number of points $s \in \mathfrak{a}_\square$ on the real and on the imaginary axis we get the same result. As the number of hyperbolas is finite, the number of all points on hyperbolas, on the real and on the imaginary axis is as well $O(x^{1/2})$. □

We are now prepared to state our main theorem:

**Theorem 4 a** *Let $\mathcal{O}^1$ be an arithmetic quaternion group in $\mathrm{SL}(2,\mathbb{C})$, whose set of traces $\mathrm{tr}\,\mathcal{O}^1$ is invariant under complex conjugation. Then the number of distinct real lengths of closed geodesics is given by*

$$\text{(56)} \qquad \mathcal{N}^r(l) = \frac{2^{2d-5}\pi}{|D_a|^{1/2}}\,\mathrm{e}^l - \frac{1}{4}\mathcal{G}(\mathrm{e}^l) + O(\mathrm{e}^{l(1-1/d)}), \qquad l \to \infty.$$

PROOF: Following lemma 3 and lemma 4, we have

$$\text{(57)} \qquad \mathcal{N}^r(l) = \frac{1}{2}\mathcal{N}^c(l) + O(\mathrm{e}^{l/2}), \qquad l \to \infty.$$

□

Again, much better remainder estimates are possible in the non-compact case:

**Theorem 4 b** *Let $\mathcal{O}^1 = \mathrm{SL}(2,\mathfrak{o}_K)$, $K = \mathbb{Q}(\sqrt{-D})$, $D \in \mathbb{N}$ square-free. Then the number of distinct real lengths of closed geodesics is given by*

(58)
$$\mathcal{N}^r(l) = \begin{cases} \dfrac{\pi}{4\sqrt{D}}\,\mathrm{e}^l + \dfrac{1}{2}\left(1 + \dfrac{1}{\sqrt{D}}\right)\mathrm{e}^{l/2} + O(\mathcal{R}_D(\mathrm{e}^l)), & \text{if } D \equiv 1, 2 \mod 4 \\[2mm] \dfrac{\pi}{2\sqrt{D}}\,\mathrm{e}^l + \dfrac{1}{2}\left(1 + \dfrac{1}{\sqrt{D}}\right)\mathrm{e}^{l/2} + O(\mathcal{R}_D(\mathrm{e}^l)), & \text{if } D \equiv 3 \mod 4, \end{cases}$$

*where $\mathcal{R}_D(x)$ denotes the remainder function of the classical circle problem for a lattice corresponding to the geometric image of $\mathfrak{o}_K$.*

PROOF: Consider first $D \equiv 1, 2 \mod 4$. Then an integer of $\mathbb{Q}(\sqrt{-D})$ is of the form $s = m + n\sqrt{-D}$, with $m, n \in \mathbb{Z}$. This means $\mathfrak{o}_K = \mathbb{Z}[\sqrt{-D}]$, hence $D_a = -4D$. As $y_\pm$ is an algebraic integer (see above) and rational, $y_\pm$ is clearly in $\mathbb{Z}$. Let us have a look at the lattice points of $\mathbb{Z}[\sqrt{-D}]$ on the hyperbolas

(59) $$e(-1): \quad m^2 - \frac{1}{3} D n^2 = 1$$

$$e(0): \quad m^2 - D n^2 = 2$$

$$e(1): \quad m^2 - 3D n^2 = 3.$$

These Diophantine equations admit an explicit solution, see, e.g., [26]. It follows easily that the number of integer solutions inside the ellipse $e(x)$ is of order $O(\log x)$, $x \to \infty$, hence negligible.

The number of points on the real axis inside $e(x)$, $x > 2$, is clearly

(60) $$2\sqrt{x+2} + O(1) = 2\sqrt{x} + O(1), \quad x \to \infty,$$

while the number of points on the imaginary axis is

(61) $$\frac{2}{\sqrt{D}}\sqrt{x+2} + O(1) = \frac{2}{\sqrt{D}}\sqrt{x} + O(1), \quad x \to \infty.$$

What ellipses $e(y)$ intersect both axes at integer points? Suppose $s_1 = m_1 \in \mathbb{Z}$ is on $e(y)$, i.e., $m_1 = \pm\sqrt{y+2} \in \mathbb{Z}$. If $s_2 = n\sqrt{-D}$ is another point on $e(y)$, then $n_2 = \pm\sqrt{(y-2)/D} = \pm\sqrt{(m_1^2 - 4)/D} \in \mathbb{Z}$. Thus $m_1$ and $n_2$ necessarily satisfy

(62) $$m_1^2 - D n_2^2 = 4.$$

The number of integer solutions of (62) inside the ellipse $e(x)$ is of order $O(\log x)$, see above. Therefore the number of all lattice points $s \in \mathbb{Z}[\sqrt{-D}]$ on the ellipses $\{e(y) \mid 2 < y \leq x,\ y \notin \mathbb{Q}\}$ adds up to

(63) $$\mathcal{E}(x) - 2\left(1 + \frac{1}{\sqrt{D}}\right)\sqrt{x} + O(\log x), \quad x \to \infty,$$

while the number of lattice points on the set of ellipses $\{e(y) \mid 2 < y \leq x,\ y \in \mathbb{Q}\}$ yields altogether

(64) $$2\left(1 + \frac{1}{\sqrt{D}}\right)\sqrt{x} + O(\log x), \quad x \to \infty.$$

There are four points on ellipses $e(y)$ with irrational $y$, compare lemma 3, and – up to $O(\log x)$ exceptions – two points on ellipses $e(y)$ with rational $y$. This proves the first case of the theorem.

Now consider $D \equiv 3 \mod 4$. Here we have $\mathfrak{o}_K = \mathbb{Z}[\frac{1+\sqrt{-D}}{2}]$, hence $D_a = -D$. For the remainder estimates, replace the lattice $\mathbb{Z}[\frac{1+\sqrt{-D}}{2}]$ by $\{m/2 + n/2\sqrt{-D} \mid m, n \in \mathbb{Z}\}$ to obtain an equivalent situation as in the first case. $\square$

Together with (5) the mean multiplicities in length spectra of three-orbifolds associated with arithmetic quaternion groups $\mathcal{O}^1$ are now asymptotically given by

$$\langle g(l) \rangle \sim c \, \frac{e^l}{l}, \qquad l \to \infty, \tag{65}$$

with

$$c = \left( \frac{2^{2d-5}\pi}{|D_a|^{1/2}} - \frac{\kappa}{4} \right)^{-1}, \tag{66}$$

where $\kappa$ is small compared to $2^{2d-3}\pi|D_a|^{-1/2}$ or even zero, if conjecture 1 holds. Recall that $\kappa = 0$, if $\mathcal{O}^1 = \mathrm{SL}(2, \mathfrak{o}_K)$.

In order to make a statement on arbitrary arithmetic lattices, it follows from the obvious bound $\mathcal{N}^r(l) \leq \mathcal{N}^c(l)$ and from the estimate (48) that there is a constant $C \geq 0$ such that

$$\lim_{l \to \infty} \frac{e^l/l}{\langle g(l) \rangle} = C, \tag{67}$$

which gives an asymptotic lower bound for the multiplicities.

# VI  Examples

The most simple arithmetic lattice in $\mathrm{SL}(2,\mathbb{C})$, one can think of, is the Picard group $\mathrm{SL}(2, \mathbb{Z}[i])$. It is generated by one half-turn and two parabolic translations,

$$\begin{pmatrix} 0 & -1 \\ 1 & 0 \end{pmatrix}, \quad \begin{pmatrix} 1 & 1 \\ 0 & 1 \end{pmatrix}, \quad \text{and} \quad \begin{pmatrix} 1 & i \\ 0 & 1 \end{pmatrix}.$$

A fundamental cell of $SL(2, \mathbb{Z}[i])$ is given by

(68) $$\mathcal{F}_\Gamma = \left\{ x \in \mathfrak{H}_3 \mid 1 \leq |x|, \ -1/2 \leq -x_2 \leq x_1 \leq 1/2 \right\},$$

see figure 1. According to lemma 1 and theorem 3, we obtain for the number of distinct complex lengths

(69) $$\mathcal{N}^c(l) = \frac{\pi}{2} e^l + O(e^{(23/73+\epsilon)l}), \quad \epsilon > 0, \quad l \to \infty,$$

as $D_\mathfrak{a} = -4$. The number of distinct real lengths reads

(70) $$\mathcal{N}^r(l) = \frac{\pi}{4} e^l + e^{l/2} + O(e^{(23/73+\epsilon)l}), \quad \epsilon > 0, \quad l \to \infty,$$

compare theorem 4 b.

Another interesting example of an arithmetic quaternion group is the cocompact lattice $T_2^+$, which is the subgroup of all orientation-preserving elements in the tetrahedral reflection group having the Coxeter diagram [27]

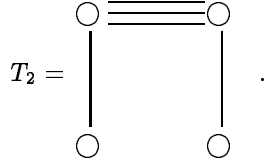

$$T_2 = \quad .$$

Each vertex $\bigcirc$ describes a face of the tetrahedron. The vertices are connected by graphs consisting of $0, 1, 2, \ldots$ lines, which represent an edge with dihedral angle $\pi/2, \pi/3, \pi/4, \ldots$, respectively. $T_2$ is therefore generated by the reflections at the faces of the hyperbolic tetrahedron $ABCD$ with dihedral angles

$$\angle BC = \pi/2, \quad \angle CA = \pi/2, \quad \angle AB = \pi/3,$$

$$\angle DA = \pi/2, \quad \angle DB = \pi/5, \quad \angle DC = \pi/3,$$

compare figure 2. $T_2^+$ is of index two in $T_2$ and generated by half-turns through the axes $BC$ and $CA$, and a $2\pi/3$-turn through $AB$. If we embed the tetrahedron into the upper half space $\mathfrak{H}_3$ like in figure 2, the generators will be represented by the SL(2, $\mathbb{C}$)-matrices

$$\begin{pmatrix} 0 & -\frac{1}{2}(\sqrt{3}-\mathrm{i}) \\ \frac{1}{2}(\sqrt{3}+\mathrm{i}) & 0 \end{pmatrix}, \quad \begin{pmatrix} 0 & \mathrm{i} \\ \mathrm{i} & 0 \end{pmatrix},$$

and

$$\begin{pmatrix} \frac{1}{2}+\frac{1}{6}\sqrt{-3+2\sqrt{5}}\sqrt{3} & -\frac{1}{6}\left(1+\sqrt{5}\right)\sqrt{3} \\ \frac{1}{6}\left(1+\sqrt{5}\right)\sqrt{3} & \frac{1}{2}-\frac{1}{6}\sqrt{-3+2\sqrt{5}}\sqrt{3} \end{pmatrix}.$$

It is worth mentioning that $T_2^+$ is a subgroup of index two in the group $\Gamma$ having the smallest covolume of all arithmetic Kleinian groups. This means, $\Gamma\backslash\mathfrak{H}_3$ is the *smallest* orientable arithmetic hyperbolic orbifold, see [28].

Let us turn to the arithmetic description of $T_2^+$. It can be viewed as the group $\mathcal{O}^1$ of units with norm one of an arbitrary maximal order in the quaternion algebra $A$, defined over the number field $K = \mathbb{Q}\left(\sqrt{3-2\sqrt{5}}\right)$ by the relations $\omega^2 = -1$, $\Omega^2 = -3$, see [29]. tr $T_2^+$ is invariant under complex conjugation, thus theorem 4 applies. To determine $D_\mathfrak{a}$, we make use of

**Lemma 5** *Let $\mathcal{O}^1$ be an arithmetic quaternion group with trace field $K$. If $\mathcal{O}^1$ contains a rotation through $2\pi/3$, then tr $\mathcal{O} = \mathfrak{o}_K$.*

PROOF: The trace of a rotation through $2\pi/3$ is one. From $1 \in \operatorname{tr} \mathcal{O}$ and from the fact that tr $\mathcal{O}$ is an ideal in $\mathfrak{o}_K$, the lemma immediately follows. □

We conclude $\mathfrak{a} = \mathfrak{o}_K$, hence $D_\mathfrak{a}$ is the discriminant of the field $K = \mathbb{Q}\left(\sqrt{3-2\sqrt{5}}\right)$, which is $D_K = -275$. The degree of $K$ clearly is $d = 4$.

In the same manner the subgroups $T_i^+$ of orientation-preserving elements in the Coxeter groups can be defined as arithmetic quaternion groups $\mathcal{O}^1$. The corresponding quaternion algebras may be looked up in [29]. It follows from the Coxeter diagrams that every group $T_i$ contains a $2\pi/3$-turn, hence again $D_\mathfrak{a} = D_K$. For $T_1$, $T_4$, $T_5$, $T_7$, $T_9$ we have $K = \mathbb{Q}\left(\sqrt{(1-\sqrt{5})/2}\right)$, $\mathbb{Q}\left(\sqrt{-1-2\sqrt{5}}\right)$, $\mathbb{Q}\left(\sqrt{-1-2\sqrt{2}}\right)$, $\mathbb{Q}\left(\sqrt{(-1-5\sqrt{5})/2}\right)$, $\mathbb{Q}\left(\sqrt{-5-4\sqrt{5}}\right)$, thus $D_K = -400, -475, -448, -775, -1375$, respectively, and each time $d = 4$.

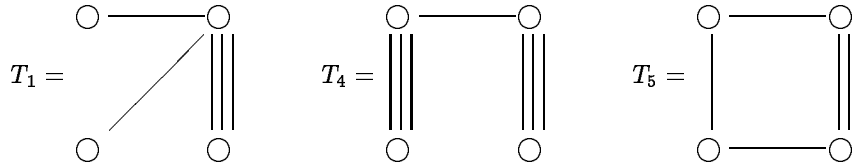

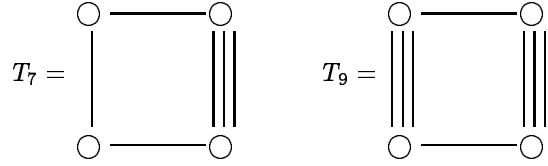

## VII  Remarks on arithmetic two-orbifolds

Hyperbolic two-orbifolds may be represented as the quotient $\Gamma\backslash\mathfrak{H}_2$, where the upper half plane $\mathfrak{H}_2$ can be viewed as a plane in $\mathfrak{H}_3$, e.g., by setting $x_2 = 0$. The group of isometries of $\mathfrak{H}_2$ is then the subgroup of $\mathrm{Iso}^+ \mathfrak{H}_3$ leaving the plane invariant, which is

(71) $$\mathrm{Iso}\,\mathfrak{H}_2 \simeq \mathrm{PSL}(2,\mathbb{R}) \cup \mathrm{PSL}(2,\mathbb{R})\sigma,$$

with

$$\sigma = \begin{pmatrix} i & 0 \\ 0 & -i \end{pmatrix}.$$

Note that $\mathrm{PSL}(2,\mathbb{R}) \cup \mathrm{PSL}(2,\mathbb{R})\sigma$ is isomorphic to $\mathrm{PGL}(2,\mathbb{R})$, but $\mathrm{PSL}(2,\mathbb{C})$ is isomorphic to $\mathrm{PGL}(2,\mathbb{C})$.

Restricting the action to $\mathfrak{H}_2$, one sees that $\mathrm{PSL}(2,\mathbb{R})$ preserves and $\mathrm{PSL}(2,\mathbb{R})\sigma$ reverses orientation ($\sigma$ is now a reflection at a line). Discrete subgroups of $\mathrm{SL}(2,\mathbb{R})$ are called *Fuchsian groups*. An arithmetic group in $\mathrm{SL}(2,\mathbb{R}) \cup \mathrm{SL}(2,\mathbb{R})\sigma$ is defined via a quaternion algebra $A$ over a totally real algebraic number field $K$ such that $A$ splits exactly at one place of $K$, i.e.,

(72) $$A \otimes_\mathbb{Q} \mathbb{R} \simeq \mathrm{M}(2,\mathbb{R}) \oplus \mathbb{H} \oplus \ldots \oplus \mathbb{H}.$$

If $\rho_1$ denotes the projection onto the first summand, restricted to $A$, and $\mathcal{O}$ is an order in $A$, then every group commensurable with $\rho_1(\mathcal{O}^1)$ (or briefly $\mathcal{O}^1$) will be called an arithmetic group.

The characterization of arithmetic groups in $\mathrm{SL}(2, \mathbb{R})$, given by Takeuchi [22], is similar to the one for groups in $\mathrm{SL}(2, \mathbb{C})$, see theorems 1 and 2.

As there are no loxodromic elements in Iso $\mathfrak{H}_2$, the relation between the traces of hyperbolic (resp. inverse hyperbolic) transformations $\gamma \in \mathrm{SL}(2, \mathbb{R}) \cup \mathrm{SL}(2, \mathbb{R})\sigma$ and lengths of closed geodesics is rather simple:

$$(73) \qquad \operatorname{tr} \gamma = \pm 2 \cosh(l_\gamma/2) \qquad (\text{resp. } \pm 2\,\mathrm{i} \sinh(l_\gamma/2)),$$

compare (11), while for elliptic (resp. inverse elliptic elements)

$$(74) \qquad \operatorname{tr} \gamma = \pm 2 \cos(\phi_\gamma/2) \qquad (\text{resp. } \pm 2\,\mathrm{i} \sin(\phi_\gamma/2)).$$

Now consider the trace field $\mathbb{Q}(\operatorname{tr} \mathcal{O}^1)$ of a unit group $\mathcal{O}^1$. It coincides with $K$. Let $K$ have degree $d$ over $\mathbb{Q}$ and denote by $\phi_j$, $j = 1, ..., d$, the embeddings of $K$ into $\mathbb{R}$. Further we suppose that $A$ splits at the first place and consequently define

$$(75) \qquad \mathfrak{a}_\square := \{ s \in \mathfrak{a} \mid -2 \leq \phi_j(s) \leq 2,\ j = 2, \ldots, d \},$$

with the ideal $\mathfrak{a} = \operatorname{tr} \mathcal{O}$. As above, let

$$\mathfrak{g} = \mathfrak{a}_\square - \operatorname{tr} \mathcal{O}^1$$

denote the set of gaps, and

$$(76) \qquad \mathcal{G}(x) := \#\{ s \in \mathfrak{g} \mid |s| \leq x \}.$$

Bolte shows

**Theorem 5** [9] *Let $\mathcal{O}^1$ be an arithmetic quaternion group in $\mathrm{SL}(2, \mathbb{R})$. Then the number of distinct lengths of closed geodesics is given by*

$$(77) \qquad \mathcal{N}^r(l) = \frac{2^{2d-2}}{|D_\mathfrak{a}|^{1/2}}\, \mathrm{e}^{l/2} - \frac{1}{2} \mathcal{G}(\mathrm{e}^{l/2}) + O(\mathrm{e}^{l(1-1/d)/2}), \qquad l \to \infty.$$

(Note that the connection between our ideal $\mathfrak{a}$ and the module $\mathcal{M}$ used in [9] is $\mathfrak{a} = 2\mathcal{M}$, hence $D_{\mathfrak{a}} = 2^{2d} D_{\mathcal{M}}$.) In analogy to conjecture 1 we formulate

**Conjecture 2** *Let $\mathcal{O}^1$ be an arithmetic quaternion group in $\mathrm{SL}(2,\mathbb{R})$. Then the number of gaps in the length spectrum up to length $l = 2\log x$ is given by*

$$(78) \qquad \mathcal{G}(x) = \kappa\, x + o(x), \qquad x \to \infty,$$

*where $\kappa \geq 0$ is a constant depending only on $\mathcal{O}^1$, and small compared to*

$$\frac{2^{2d-1}}{|D_{\mathfrak{a}}|^{1/2}}.$$

This is a little more cautious than $\kappa = 0$ assumed in [9].

To derive the number of distinct lengths $\mathcal{N}^r(l;\Gamma)$ for general arithmetic Fuchsian groups $\Gamma$, Bolte claims that if $\Gamma$ and $\Gamma'$ are commensurable, then

$$(79) \qquad I'\, \mathcal{N}^r(l;\Gamma) \sim I\, \mathcal{N}^r(l;\Gamma'), \qquad l \to \infty,$$

where $I$ is the index of $\Gamma$ over $\Gamma \cap \Gamma'$, and $I'$ the index of $\Gamma'$ over $\Gamma \cap \Gamma'$. This is, however, not true in general. We will give a simple counter example: To stand on solid ground, let us choose the modular group $\Gamma = \mathrm{SL}(2,\mathbb{Z})$, where no gaps occur (compare lemma 1). $\mathrm{SL}(2,\mathbb{Z})$ is the group of units with norm one in the quaternion order $\mathrm{M}(2,\mathbb{Z})$. Clearly $\mathfrak{a} = \mathbb{Z}$ with $D_{\mathfrak{a}} = 1$, $d = 1$, therefore

$$(80) \qquad \mathcal{N}^r(l;\Gamma) = \mathrm{e}^{l/2} + O(1), \qquad l \to \infty.$$

Now consider the principal congruence group

$$(81)$$
$$\Gamma(N) = \left\{ \begin{pmatrix} a & b \\ c & d \end{pmatrix} \in \mathrm{SL}(2,\mathbb{Z}) \,\bigg|\, \begin{pmatrix} a & b \\ c & d \end{pmatrix} \equiv \begin{pmatrix} 1 & 0 \\ 0 & 1 \end{pmatrix} \mod N \right\},$$

then $\Gamma' = \Gamma(N) \cup \Gamma(N)(-1)$ is of finite index in $\mathrm{SL}(2,\mathbb{Z})$,

$$(82) \qquad I = \begin{cases} 6, & \text{if } N = 2 \\ \dfrac{N^3}{2} \displaystyle\prod_{p|N} \left(1 - \dfrac{1}{p^2}\right), & \text{if } N > 2, \end{cases} \qquad I' = 1,$$

see [30, 31]. Following lemma 1, $\operatorname{tr}\Gamma(N) = \{s \in \mathbb{Z} \mid s \equiv 2 \mod N^2\}$, hence

$$(83) \quad \mathcal{N}^\tau(l;\Gamma') = \mathcal{N}^\tau(l;\Gamma(N)) = \begin{cases} \dfrac{1}{4} e^{l/2} + O(1), & \text{if } N = 2 \\ \dfrac{2}{N^2} e^{l/2} + O(1), & \text{if } N > 2, \end{cases}$$

which contradicts (79).

**Acknowledgements:** I am very grateful to Professor F. Steiner and Dr. Jens Bolte for introducing me into the theory of quantum chaos.

# Figures

Figure 1: Fundamental cell of the Picard group. Boundary points are identified by half-turns through the axes $DM_{AB}$, $DM_{BC}$, $DM_{CA}$ and $BM_{CA}$, where $M_{AB}$ denotes the midpoint between $A$ and $B$, $M_{BC}$ the one between $B$ and $C$, etc.

Figure 2: A Fundamental cell of the subgroup $T_2^+$ of orientation-preserving elements in $T_2$ is the tetrahedron $ABD'D$. Boundary points are identified by the generators of $T_2^+$: two half-turns through the axes $BC$ and $CA$, and a $2\pi/3$-turn through $AB$. The tetrahedron $ABCD$ is the fundamental cell of $T_2$, generated by the reflections at the faces of $ABCD$.

33